\documentstyle[epsfig]{mn}
\newif\ifAMStwofonts
\def\ltsima{$\; \buildrel < \over \sim \;$}
\def\simlt{\lower.5ex\hbox{\ltsima}}
\def\gtsima{$\; \buildrel > \over \sim \;$}
\def\simgt{\lower.5ex\hbox{\gtsima}}
\ifoldfss
  \ifCUPmtlplainloaded \else
    \NewTextAlphabet{textbfit} {cmbxti10} {}
    \NewTextAlphabet{textbfss} {cmssbx10} {}
    \NewMathAlphabet{mathbfit} {cmbxti10} {} 
    \NewMathAlphabet{mathbfss} {cmssbx10} {} 
  \fi
  \ifAMStwofonts
    \ifCUPmtlplainloaded \else
      \NewSymbolFont{upmath} {eurm10}
      \NewSymbolFont{AMSa} {msam10}
      \NewMathSymbol{\upi}     {0}{upmath}{19}
      \NewMathSymbol{\umu}     {0}{upmath}{16}
      \NewMathSymbol{\upartial}{0}{upmath}{40}
      \NewMathSymbol{\leqslant}{3}{AMSa}{36}
      \NewMathSymbol{\geqslant}{3}{AMSa}{3E}

    \fi
  \fi
\fi 

\ifnfssone
  \newmathalphabet{\mathit}
  \addtoversion{normal}{\mathit}{cmr}{m}{it}
  \addtoversion{bold}{\mathit}{cmr}{bx}{it}
  \newmathalphabet{\mathbfit} 
  \addtoversion{normal}{\mathbfit}{cmr}{bx}{it}
  \addtoversion{bold}{\mathbfit}{cmr}{bx}{it}
  \newmathalphabet{\mathbfss} 
  \addtoversion{normal}{\mathbfss}{cmss}{bx}{n}
  \addtoversion{bold}{\mathbfss}{cmss}{bx}{n}
  \ifAMStwofonts
    \ifCUPmtlplainloaded \else
      \UseAMStwoboldmath
      \makeatletter
      \new@mathgroup\upmath@group
      \define@mathgroup\mv@normal\upmath@group{eur}{m}{n}
      \define@mathgroup\mv@bold\upmath@group{eur}{b}{n}
      \edef\UPM{\hexnumber\upmath@group}
      \new@mathgroup\amsa@group
      \define@mathgroup\mv@normal\amsa@group{msa}{m}{n}
      \define@mathgroup\mv@bold\amsa@group{msa}{m}{n}
      \edef\AMSa{\hexnumber\amsa@group}
      \makeatother
      \mathchardef\upi="0\UPM19
      \mathchardef\umu="0\UPM16
      \mathchardef\upartial="0\UPM40
      \mathchardef\leqslant="3\AMSa36
      \mathchardef\geqslant="3\AMSa3E
    \fi
  \fi
\fi 

\ifnfsstwo
  \DeclareMathAlphabet{\mathbfit}{OT1}{cmr}{bx}{it}
  \SetMathAlphabet\mathbfit{bold}{OT1}{cmr}{bx}{it}
  \DeclareMathAlphabet{\mathbfss}{OT1}{cmss}{bx}{n}
  \SetMathAlphabet\mathbfss{bold}{OT1}{cmss}{bx}{n}
  \ifAMStwofonts
    \ifCUPmtlplainloaded \else
      \DeclareSymbolFont{UPM}{U}{eur}{m}{n}
      \SetSymbolFont{UPM}{bold}{U}{eur}{b}{n}
      \DeclareSymbolFont{AMSa}{U}{msa}{m}{n}
      \DeclareMathSymbol{\upi}{0}{UPM}{"19}
      \DeclareMathSymbol{\umu}{0}{UPM}{"16}
      \DeclareMathSymbol{\upartial}{0}{UPM}{"40}
      \DeclareMathSymbol{\leqslant}{3}{AMSa}{"36}
      \DeclareMathSymbol{\geqslant}{3}{AMSa}{"3E}
    \fi
  \fi
\fi 

\ifCUPmtlplainloaded \else
  \ifAMStwofonts \else 
    \def\upi{\pi}
    \def\umu{\mu}
    \def\upartial{\partial}
  \fi
\fi

\title[The relationship between the X-ray variability and the central black
hole mass]{The relationship between the X-ray variability and the central
black hole mass}
\author[Y. Lu \& Q. Yu]
       {Youjun Lu$^{1,2}$\thanks{Current Address: Princeton University Observatory, Princeton, NJ 08544-1001, USA; Email: lyj@astro.princeton.edu}
       and Qingjuan Yu$^3$\thanks{Email: yqj@astro.princeton.edu}\\
       $\ ^1$ Centre for Astrophysics, University of Science
        and Technology of China, Hefei Anhui 230026, P.\ R.\ China.\\
       $\ ^2$ National Astronomical Observatories, Chinese Academy of Sciences\\
       $\ ^3$ Princeton University Observatory, Princeton, NJ 08544-1001, USA. 
       \\}

\pubyear{2001}

\begin{document}
\maketitle

\label{firstpage}

\begin{abstract}
We assembled a sample of Seyfert 1 galaxies, QSOs and Low-Luminosity
Active Galactic Nuclei (LLAGNs) observed by ASCA, whose central
black hole masses have been measured. We found that the X-ray
variability (which is quantified by the ``excess variance'' $\sigma^2_{
\rm rms}$) is significantly anti-correlated with
the central black hole mass, and there likely exists a linear relationship $
\sigma^2_{\rm rms}\propto M^{-1}_{\rm bh}$. It can be interpreted that 
the short time-scale X-ray variability is caused by some global 
coherent variations in the X-ray emission region which is scaled by
the size of the central black hole. Hence, the central black hole mass
is the driving parameter of the previously established relation between
X-ray variability and luminosity. Our findings favor the hypothesis
that the Narrow Line Seyfert 1 galaxies and QSOs harbor smaller black
holes than the broad line objects, and can also easily explain the
observational fact that high redshift QSOs have greater variability
than local AGNs at a given luminosity. 
Further investigations are needed to confirm our findings, and a large
sample X-ray variability investigation can give constraints on the
physical mechanism and evolution of AGNs.
\end{abstract}

\begin{keywords}
galaxies: active -- galaxies: nuclei -- galaxies: Seyfert -- X-rays: galaxies.
\end{keywords}

\section{Introduction}

X-ray variability has long been extensively studied for Active
Galactic Nuclei (AGN) (see Mushotzky, Done \& Pounds \shortcite{mdp}
and references
therein ). In general, the X-ray flux from AGNs exhibits not only
long-term variability, but also very rapid variability on
time-scales of less than thousands of seconds. This short time-scale
variability indicates that X-ray emission most likely originates
in the innermost regions of an AGN, and thus can help unravel the
basic parameters of the central engine of AGNs ( e.g. mass, accretion
rate, geometry and radiation mechanisms). 

Previous X-ray observations have revealed that the source flux
doubling time-scale is significantly correlated with luminosity 
\cite{bm}. Using the long time uninterrupted observations
of EXOSAT, Lawrence \& Papadakis \shortcite{lp} and Green, McHardy \& Lehto
\shortcite{gml} found that the
power density spectra (hereafter PDS) of AGNs are consistent with a
single form, but the amplitudes are strongly anti-correlated with the X-ray
luminosities. These results have been confirmed by ASCA data. Nandra et al.
\shortcite{nandra} first used the ``excess variance'' to quantify the X-ray
variability of AGNs, and found a significant anti-correlation between
X-ray variability and luminosity for an ASCA sample of AGNs with
predominantly broad emission line objects. Turner et al. \shortcite{turner} and
Leighly \shortcite{leighly} further extended this investigation to
the Seyfert 1 galaxies and QSOs with narrow lines (hereafter NLS1),
and found that it introduces
larger scatter into the established correlation between X-ray
variability and luminosity. Moreover, Turner et al. \shortcite{turner} found
a significant correlation between hard X-ray variability and FWHM of
H$\beta$, and this introduces a connection between the X-ray variability
and the ``Eigenvector 1'' parameters which is proposed to be related
with the fundamental parameters of the central engine \cite{bg}.

Some possible origins of the correlation between the ``excess variance''
and the luminosity in X-ray band have been discussed by many authors
\cite{lp,ba,nandra,leighly,ac}. The most obvious one might be
that the correlation is related to the source size. One might expect that
the characteristic time scales or the ``excess variance'' should scale or
relate with the central black hole mass \cite{en}. However, the distribution
of the other parameters (probably the inclination, the accretion rate and the
absorption properties) in the sample can also make the correlation. In fact,
Bao \& Abramowicz \shortcite{ba} suggested that the
X-ray variability is produced by bright-spots on a rotating accretion disk,
and that the correlation is primarily a projection effect. In their model,
X-ray variability was enhanced due to the increased relativistic effects
with the change of inclination from a face-on to an edge-on disk.
However, this model conflicts with the unified model of AGNs in which
typical Seyfert 1 galaxies are observed close to face-on. Recently, Abrassart
\& Czerny \shortcite{ac} proposed an alternative explanation for the
variability of AGNs based on a cloud model of accretion onto a black hole.
The small random rearrangement of the cloud distribution can happen on 
time-scales on the order
of $10^2-10^6$s and may lead to relatively high amplitude variability
in X-ray based on the large mean covering factor. However, the number
of the clouds and the mean covering factor may be determined by the
accretion rate and the central black hole mass, and thus the 
X-ray variability may be related to some fundamental parameters of the
central engine in this regime.

Due to the endeavors of the past decade, the pursuing of measuring the
central black hole mass in the galaxies nuclei has resulted in fruitful
achievements.  
First, the central massive dark objects$-$possibly the supermassive black
hole$-$have been measured in the centre of some nearby galaxies based on high
spatial resolution observations of stellar dynamics \cite{kr,magorr}.
A tight correlation has been found between the central black hole mass
and the stellar velocity dispersion which links the inner nuclei with the
large scale galactic environment \cite{fm,g2000a}.
Secondly, reverberation mapping data can give a reliable estimation of 
the central black hole mass in AGNs (Peterson \& Wandel 1998; Wandel,
Peterson \& Malkan 1999, hereafter WPM; Kaspi et al. 2000, hereafter KSNMGJ),
which cannot be determined by the stellar dynamics due to the bright nuclei.
The interesting thing is that the relation between the mass of black hole
and the bulge gravitational potential also exists in AGNs (Gebhardt et al.
2000b; Nelson 2000). This consistency may further support that the
reveberation mapping method is reliable to determine the black hole mass
as the stellar dynamics.
All these measurements have somewhat distangled the fundamental parameters
of the central engine, and allow us to study the relations between the
``excess variance'' and the fundamental parameters (e.g. the black hole mass
and the accretion rate) , and thus shed new light on our understanding
of the central engine.

In this paper, we investigate the links between the ``excess variance''
and the fundamental parameters of the central engine, i.e. the central
black hole mass. The sample and data reduction
are presented in \S 2; the statistical analysis and results are presented
in \S 3 and discussed in \S 4; and finally, the conclusions are summarized
in \S 5. 

\section{Sample and Data Reduction}

  The masses of the central black holes for a bunch of Seyferts and QSOs
have been measured using reverberation mapping data (WPM, Ho 1998, and 
KSNMJG), based on the assumption that the line-emission material is
gravitationally bound and hence has a near-Keplerian velocity dispersion.
This assumption is supported by the recent investigation on the Seyfert
1 galaxy NGC5548 which demonstrates that the kinematics of the broad line
region is Keplerian \cite{pw}. There are two kinds of methods
to estimate a typical velocity and hence the central black hole mass:
one is to measure the Full Width at Half Maximum (FWHM) of each line in
all spectra and calculate the mean FWHM, the other uses the the rms
spectrum to compute the FWHM of the lines (WMP and KSNMJG); the
Virial mass is then determined to be $M_{\rm bh}$(mean) or/and $M_{\rm bh}$(rms)
using FWHM(mean) or/and FWHM(rms) together with the time lags of the lines 
corresponding to the variance of the ionizing continuum (see WPM and KSNMJG
for details).

We searched in the public ASCA archive up to Oct. 1999 for the sample
objects of KSNMJG and Ho \shortcite{ho} (the objects in WPM sample are
included in the KSNMJG sample), of which the masses of the central black
holes have been measured using the reverberation data, and found
that 24 objects have been observed by ASCA. Except two objects,
PG1411+442 and PG1700+518, which cannot meet the criteria to do the timing
analysis (see the following data processing procedures), all the others 
are listed in table~\ref{tab1} with the measured Virial reverberation mass
(both $M_{\rm bh}$(mean) and $M_{\rm bh}$(rms); KSNMJG and Ho 1998).

\begin{table*}
\begin{minipage}{90mm}
\caption{Variability properties and the central black hole masses for AGNs:
The central black hole masses are adopted from Table~5 in KSNMJG except for
those objects labelled with $^a$, from which the central black hole mass are
adopted from Ho (1998). The ``excess variance'' are adopted from Turner et 
al. (1999) except for those objects labelled with $~b$, which are measured 
in this work. }
\label{tab1}
\begin{tabular}{lcccc} \hline \hline
Name        & Sequence  &  $\sigma^2_{\rm rms}(10^{-3})$ & $\frac{M_{\rm bh}({\rm mean})}{10^7M_{\sun}}$  &  $\frac{M_{\rm bh}({\rm rms})}{10^7M_{\sun}}$ \\  \hline
3C120       & 71014000  &  2.95$\pm$0.71  & 2.3$^{+1.5}_{-1.1}$    &
3.0$^{+1.9}_{-1.4}$   \\
3C390.3     & 73082000  &  0.00$\pm$0.86  & 34$^{+11}_{-13}$       &
37$^{+12}_{-14}$      \\
            & 73082010  &  0.00$\pm$0.52  & $\cdots \cdots$        &
$\cdots \cdots$       \\
NGC3227     & 70013000  &  51.7$\pm$14.0  & 3.9$^{+2.1}_{-3.9}$    &
4.9$^{+2.6}_{-4.9}$   \\
            & 73068000  &  20.3$\pm$4.7   & $\cdots \cdots$        &
$\cdots \cdots$       \\
AKN120      & 72000000  &  1.31$\pm$0.35  & 18.4$^{+3.9}_{-4.3}$   &
18.7$^{+4.0}_{-4.4}$  \\
MARK335     & 71010000  &  4.86$\pm$1.59  & 0.63$^{+0.23}_{-0.13}$ &
0.38$^{+0.14}_{-0.10}$ \\
Fairall9    & 71027000  &  0.92$\pm$0.59  & 8.0$^{+2.4}_{-4.1}$    &
8.3$^{+2.5}_{-4.3}$    \\
            & 73011000  &  0.82$\pm$0.45  & $\cdots \cdots$        &
$\cdots \cdots$        \\
            & 73011010  &  0.15$\pm$0.34  & $\cdots \cdots$        &
$\cdots \cdots$        \\
            & 73011020  &  1.72$\pm$0.75  & $\cdots \cdots$        &
$\cdots \cdots$        \\
            & 73011030  &  0.55$\pm$0.77  & $\cdots \cdots$        &
$\cdots \cdots$        \\
            & 73011040  &  2.76$\pm$0.79  & $\cdots \cdots$        &
$\cdots \cdots$        \\
            & 73011060  &  1.02$\pm$0.57  & $\cdots \cdots$        &
$\cdots \cdots$        \\
IC4329A     & 70005000  &  1.48$\pm$0.76  & 0.5$^{+1.3}_{-1.1}$    &
0.7$^{+1.8}_{-1.6}$    \\
MARK509     & 71013000  &  0.95$\pm$0.21  & 5.78$^{+0.68}_{-0.66}$ &
9.2$^{+1.1}_{-1.1}$    \\
            & 74024030  &  0.77$\pm$0.45  & $\cdots \cdots $       &
$\cdots \cdots$        \\
            & 74024040  &  0.11$\pm$0.28  & $\cdots \cdots $       &
$\cdots \cdots$        \\
            & 74024050  &  0.00$\pm$0.12  & $\cdots \cdots $       &
$\cdots \cdots$        \\
            & 74024060  &  0.00$\pm$0.30  & $\cdots \cdots $       &
$\cdots \cdots$        \\
            & 74024070  &  0.00$\pm$0.30  & $\cdots \cdots $       &
$\cdots \cdots$        \\
            & 74024080  &  0.74$\pm$0.37  & $\cdots \cdots $       &
$\cdots \cdots$        \\
            & 74024090  &  0.86$\pm$0.55  & $\cdots \cdots $       &
$\cdots \cdots$        \\
NGC3783     & 71041000  &  7.91$\pm$2.20  & 0.94$^{+0.92}_{-0.84}$ &
1.10$^{+1.07}_{-0.98}$ \\
            & 71041010  &  5.33$\pm$1.04  & $\cdots \cdots $       &
$\cdots \cdots$        \\
            & 74054000  &  0.92$\pm$0.36  & $\cdots \cdots $       &
$\cdots \cdots$        \\
            & 74054010  &  1.75$\pm$0.70  & $\cdots \cdots $       &
$\cdots \cdots$        \\
            & 74054020  &  2.91$\pm$1.12  & $\cdots \cdots $       &
$\cdots \cdots$        \\
            & 74054030  &  4.13$\pm$1.33  & $\cdots \cdots $       &
$\cdots \cdots$        \\
NGC4051     & 70001000  &  126.0$\pm$24.0 & 0.13$^{+0.13}_{-0.08}$ &
0.14$^{+0.15}_{-0.09}$ \\
            & 72001000  &  162.0$\pm$22.4 & $\cdots \cdots $       &
$\cdots \cdots$         \\
NGC4151     & 71019020  &  3.54$\pm$1.34  & 1.53$^{+1.06}_{-0.89}$ &
1.20$^{+0.83}_{-0.70}$  \\
            & 71019010  &  6.05$\pm$1.23  & $\cdots \cdots $       &
$\cdots \cdots$        \\
NGC5548     & 70018000  &  5.49$\pm$3.00  & 12.3$^{+2.3}_{-1.8}$   &
9.4$^{+1.7}_{-1.4}$    \\
            & 70038000  &  0.50$\pm$0.27  & $\cdots \cdots $       &
$\cdots \cdots$        \\
            & 74038010  &  1.57$\pm$0.46  & $\cdots \cdots $       &
$\cdots \cdots$        \\
            & 74038020  &  0.89$\pm$0.65  & $\cdots \cdots $       &
$\cdots \cdots$        \\
            & 74038030  &  0.26$\pm$0.39  & $\cdots \cdots $       &
$\cdots \cdots$        \\
            & 74038040  &  0.15$\pm$0.28  & $\cdots \cdots $       &
$\cdots \cdots$        \\
NGC7469     & 71028030  &  5.30$\pm$3.20  & 0.65$^{+0.64}_{-0.65}$ &
0.75$^{+0.74}_{-0.75}$ \\
MARK279$~^a$     & 72028000  &  2.80$\pm$0.63  & 4.20                   &
$\cdots \cdots$        \\
NGC4593$~^a$     & 71024000  &  21.3$\pm$3.45  & 0.81                   &
$\cdots \cdots$        \\
NGC3516$~^a$     & 71007000  &  7.31$\pm$1.21  & 2.30                   &
$\cdots \cdots$        \\
PG0804+761$~^b$  & 75058000  &  1.61$\pm$0.74  & 18.9$^{+1.9}_{-1.7}$   &
16.3$^{+1.6}_{-1.5}$    \\
PG0844+349$~^b$      & 76059000  &  10.5$\pm$3.23  & 2.16$^{+0.90}_{-0.83}$ &
2.7$^{+1.1}_{-1.0}$     \\
PG0953+415$~^b$      & 75060000  &  0.30$\pm$2.90  & 18.4$^{+2.8}_{-3.4}$   &
16.4$^{+2.5}_{3.0}$     \\
PG1211+243$~^b$  & 70025000  &  6.89$\pm$2.93  & 4.05$^{+0.96}_{-1.21}$ &
2.36$^{+0.56}_{-0.70}$  \\
MARK110$~^b$     & 73091000  &  1.41$\pm$0.47  & 0.56$^{+0.20}_{-0.21}$ &
0.77$^{+0.28}_{-0.29}$  \\
3C273$~^b$       & 70023000  &  0.46$\pm$0.13  & 55.0$^{+8.9}_{-7.9}$   &
23.5$^{+3.7}_{-3.3}$    \\
            & 10402000  &  1.30$\pm$0.25  & $\cdots \cdots$        &
$\cdots \cdots$         \\
            & 12601000  &  0.32$\pm$0.11  & $\cdots \cdots$        &
$\cdots \cdots$         \\ \hline
\end{tabular}
\end{minipage}
\end{table*}

\begin{table*}
\begin{minipage}{90mm}
\caption{The variability properties, central black hole masses for LLAGNs. }
\label{tab2}
\begin{tabular}{lccc} \hline \hline
Name        & Sequence Number & $\sigma^2_{\rm rms}(10^{-3})$ & $\frac{M_{\rm bh
}}{10^7M_{\sun}}$  \\ \hline
NGC4258     &   60000180 &  0.75 $\pm$ 1.95 & 3.60   \\
            &   64000000 &   6.41 $\pm$ 1.79 & $\cdots$ \\
            &   64000010 &   7.95 $\pm$ 1.92 & $\cdots$ \\
            &   64000020 &   5.65 $\pm$ 1.83 & $\cdots$  \\
NGC4579     &   73063000 &   6.10 $\pm$ 1.74 & 0.40    \\
M81         &   15000030 &   1.28 $\pm$ 0.45 & 0.40    \\
            &   15000050 &   4.98 $\pm$ 1.55 & $\cdots$ \\
            &   15000120 &   2.03 $\pm$ 1.89 & $\cdots$ \\
            &   15000130 &   1.71 $\pm$ 2.23 & $\cdots$ \\
\hline
\end{tabular}
\end{minipage}
\end{table*}

For consistent with the former investigations, we adopted the same method
used by Nandra et al. \shortcite{nandra} and Turner et al. \shortcite{turner}
to do the timing analysis and extract the ``excess variance'' as follows.
We used a 256-second time bin for the light curve. We combined the SIS0 and
SIS1 data to increase the signal-to-noise ratio and required all time bins
to be at least 99\% exposed. We required that the time series in our analysis
have at least 20 counts per bin and at least 20 bins in the final light curve.
The background was not subtracted from the light curves since it has been
demonstrated by Turner et al. (1999) and Leighly (1999) that in practice
the background is only a small fraction of the photons and it only has a
small effects on the value of ``excess variance'' for light curves with
SIS0+SIS1 counts rate greater than 0.5 counts per second. Even for the objects
(PG0844+349, PG0953+415 and NGC4579) which have SIS0+SIS1 counts rate less
than 0.5 counts per second in our sample, we found that the difference is not
large.  Finally, we calculated the normalized ``excess variance'', $$
\sigma^2_{\rm rms} = \frac{1}{N\mu^2}
\sum_{i=1}^N [(X_i-\mu)^2-\sigma^2_i]\ ,
$$
where $X_i$ with error $\sigma_i$ is the count rates for the i points
of the light curve with total N points, and $\mu$ is the mean value of
$X_i$. The error on $\sigma^2_{\rm rms}$ is given by $s_D/(\mu^2\sqrt{N})$,
where
$$
s^2_D=\frac{1}{N-1}
\sum_{i=1}^N {\{[(X_i-\mu)^2-\sigma^2_i]-\sigma^2_{\rm rms}\mu^2\}}^2\
$$
which is only the statistical error. For 16 out of the objects in 
table~\ref{tab1},
of which the ``excess variance'' for the combined data of two solid-state
imaging spectrometers (SISs) have been measured by Turner et al.
\shortcite{turner}, we re-calculated the ``excess variance'' as a check
and the values are almost the same as those tabulated in their paper.
Therefore, in the present paper, we adopted the values from Turner et al.
\shortcite{turner}. For the other six objects, the measured values are also
listed in table~\ref{tab1}. Note the ``excess variance'' for PG1211+143 is
almost the same as that measured by Leighly (1999) which was measured by a
slightly different method. In particular, there are multiple ASCA observations
for 3C273, and the observations which can meet the criteria are listed in
table~\ref{tab1}.

We noticed that Almaini et al. \shortcite{alm} introduced a more precise
maximum likelihood estimator of the ``excess variance'', which can be reduced
to the form of Nandra et al. \shortcite{nandra} if the measurement errors
are almost identical. All the points of any light curve in our sample
have approximately the same measurement errors, and thus the maximum likelihood
method should give almost an identical value of the ``excess variance''
with that given in table~\ref{tab1}.

Ptak et al. \shortcite{ptak} measured the X-ray variability for a sample of
LLAGNs and found that LLAGNs tend to show little or no significant short
term variability. For five objects in their sample, M81, NGC3079, NGC4258,
NGC4579 and NGC4594, the black holes are identified at the centre of
the galaxies \cite{kr,miyoshi,magorr}. In order to investigate the relations
between $\sigma^2_{\rm rms}$ and the central black hole mass, we chose them
as our sample objects and calculated $\sigma^2_{\rm rms}$ of the combined
SISs data in 0.5-10keV band using the same criteria. Unfortunately, NGC3079
and NGC4594 do not meet the criteria described above. We threw these two
objects out of our sample.\footnote{However, one may notice that in the
above five objects, NGC3079 has the smallest central black hole with mass
$1.3\times10^6M_{\sun}$, while NGC4594 has the largest one with mass
$6.5\times10^8M_{\sun}$. Using the gas imaging spectrometers (GISs) 
2-10keV data, Ptak et al. \shortcite{ptak} measured the ``excess variance''
for those objects. NGC3079 has the largest $\sigma^2_{\rm rms}$, while NGC4594
has a negative $\sigma^2_{\rm rms}$ which shows no significant variability.
This is compatible with the relation between the ``excess variance'' and the
black hole mass discussed in the following sections.} The calculated
values of $\sigma^2_{\rm rms}$ for the others are listed in table~\ref{tab2}.
Note the fact that the dominant soft components at 0.4-20keV band of LLAGNs, 
which are probably thermal and not associated with AGNs, may wash out the
variability in hard power-law components \cite{ptak}. It means we may
underestimate the ``excess variability'' for those LLAGNs.

\section{Statistical Analysis and Results}

\begin{figure}
 \centerline{\psfig{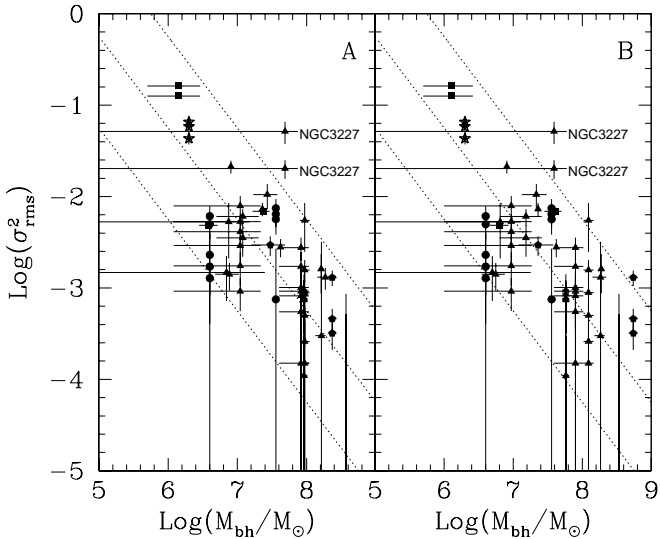}}
 \vskip 0.2cm
 \caption{The ``excess variance'' versus the central black hole mass:
$M_{\rm bh}$(rms) and $M_{\rm bh}$(mean) are used in Figure~\ref{fig-1} A and
B for those objects which have been measured in two ways. In both plots, the
filled triangles represent the radio-quiet Seyfert 1 galaxies and QSOs with
broad optical lines; the filled squares represent the Narrow Line Seyfert 1
galaxies and QSOs; the circles represent LLAGNs; the filled pentagons represent
the radio-loud objects; and the stars represent MCG-6-30-15, a special
object discussed in section 4. }
 \label{fig-1}
\end{figure}

Figure~\ref{fig-1} show $\sigma^2_{\rm rms}$ as a function of the measured mass
of the central black hole $M_{\rm bh}$. For those objects with the masses
measured in two ways, $M_{\rm bh}$(rms) and $M_{\rm bh}$(mean) are used
in Figure~\ref{fig-1} A and B, respectively. There is clearly a trend that the
objects with small central black holes exhibit large variabilities in both
plots, although with a large amount of scatter. In fact, a very tight 
correlation could not be expected because there are considerable uncertainties
in both the black hole mass determination and the $\sigma^2_{\rm rms}$
measurements. For example, the measured values of $\sigma^2_{\rm rms}$
are affected by a lot of factors, such as the length of the observations
and the total time spanned of the observations, etc. (see Leighly 1999
for details); we may also underestimate $\sigma^2_{\rm rms}$ of LLAGNs 
using the SISs data in 0.5-10keV band since the thermal soft component 
LLAGNs in 0.5-2.0keV bandpass may ``wash out'' some variability. Moreover,
significant changes are evidently observed in $\sigma^2_{\rm rms}$ 
among the X-ray observations of a single source, and we used multiple ASCA
observations for several objects which were observed frequently. However,
it is evident from Figure~\ref{fig-1} that LLAGNs and Seyferts/QSOs all follow
the same trend, although the luminosities of LLAGNs are much smaller than
those of Seyferts and QSOs. If we consider the ``wash out'' fact for LLAGNs,
the correlation should be tighter.

One may notice that NGC3227 deviates from the main trend in Figure~\ref{fig-1}. 
This object is unusual by virtue of its very flat soft X-ray spectrum 
which may be caused by a dusty warm absorber \cite{george,kf}.  
Probably its variability is enhanced by some changes in the absorber. 
In addition, Schinnerer, Eckart \& Tacconi \shortcite{set} reported that
the enclosed mass in the inner 25~pc of NGC3227 is about $2\times10^7M_{
\odot}$ based on a detection of molecular gas at a distance from nucleus 
of only $\sim15$~pc. Although this mass approximately agrees with the 
Virial mass of $3.9-4.9\times10^7M_{\sun}$ measured by using reverberation
mapping data with larger uncertainty, it may suggest that the mass of 
the central black hole is lower than the measured Virial mass. If 
the central black hole mass of NGC3227 is really lower than the estimated
Virial mass (say, by a factor of 2), NGC3227 would join the main
trend. Furthermore, we should caution that some systematic error may exist
because the adopted masses are measured by different techniques for LLAGNs
and Seyfert 1 galaxies/QSOs. This systematic error should not be very large
since the measured masses by these two methods follow the same relation 
with the galaxies bulge potential \cite{g2000b,nelson},
and the trend still remains even if we exclude those points for LLAGNs in
Figure~\ref{fig-1}.

In Figure~\ref{fig-1}, the trend seems to be the case that there is a linear
relationship, $\sigma^2_{\rm rms}\propto M^{-1}_{\rm bh}$. As we can see,
it can be represented by a line $\log\sigma^2_{\rm rms}=4.75-\log(
M_{\rm bh}/M_{\odot})$ in both Figure~\ref{fig-1} A and B. All the 
objects are localized in the region between the lines $\log\sigma^2_{\rm 
rms}=3.75-\log(M_{\rm bh}/M_{\odot})$ and 
$\log\sigma^2_{\rm rms}=5.75-\log(M_{\rm bh}/M_{\odot})$ except NGC3227.
More quantitatively, a Spearman rank test gives the correlation
coefficients of -0.70 and -0.65 for the points in Figure~\ref{fig-1} A and B,
respectively; and rejects the possibility that $\sigma^2_{\rm rms}$ and
$M_{\rm bh}$ are uncorrelated at $>$99.9\% confidence. The robust nature of
a rank test means that the significance of this correlation does not depend on
the outlying point NGC3227. If we include NGC3227 in the test, the
corresponding correlation coefficient are -0.68 and -0.64 with almost 
the same confidence for the points in Figure~\ref{fig-1} A and B, respectively.

\section{Discussions}

In the present paper, we found that the ``excess variance'' is significantly
anti-correlated with the central black hole mass for a combined
sample of Seyfert 1 galaxies, QSOs and LLAGNs. The most plausible
explanation is that the ``excess variance'' is caused by some global
coherent changes in the X-ray emitting region, and this region scales with
the size of black hole. The light curves are known to be characterized
by a steep power-law PDS ($P(f)\propto f^{-\alpha}$, where $\alpha
\sim 1.5-2$) in some AGNs, such as NGC4051, NGC3516, NGC5548 and MCG-6-30-15
\cite{lp,nch}. Assuming self similar scaling and hence a direct connection
between time scales and the size of sources, the observed ``excess variance''
can be related to the size of the central black
hole as $\sigma^2_{\rm rms} = \int_{f_1}^{f_2}P(f)df \propto f^{1-\alpha}_{1}
\propto R^{1-\alpha} \propto M^{1-\alpha}_{\rm bh}$, where $f_1\ll f_2$ and
$\alpha\neq 1$. One can readily get the observed correlation $
\sigma^2_{\rm rms} \propto M^{-1}_{\rm bh}$ illustrated in Figure~\ref{fig-1}
by assuming $\alpha \sim 2$. This fundamental relationship can
self-consistently explain the previous finding of the relationship between
$\sigma^2_{\rm rms}$ and luminosity, which has been proposed by many authors
\cite{nandra,turner,leighly,alm}.

Recently, some investigations have been performed on the relationship between
$\sigma^2_{\rm rms}$ and luminosity for a sample of LLAGNs \cite{ptak}
and a deep flux limited sample of QSOs selected from deep ROSAT survey
\cite{alm}, which greatly extend the luminosity range and redshift range.
Ptak et al. \shortcite{ptak} found that LLAGNs tend to show little or no
significant short term variability, and there is a break from the trend
of increased variability in Seyfert 1 galaxies with decreased luminosity.
They proposed that this is due to the lower accretion rates in LLAGNs. They
argued that this results in a larger characteristic size of the X-ray emission
region in LLAGNs than in Seyfert 1 galaxies because the lower accretion rate
is probably causing the accretion flow to be advection-dominated. However,
most of the X-ray emission should originate in an inner volume probably
with a radius less than 10$R_{\rm Sch}$ \cite{ptak}, which is similar to the
typical X-ray emission region size ($\sim 10R_{\rm Sch}$) of a normal Seyfert
1 galaxies. If the X-ray variability is caused by some global coherent 
oscillation for both LLAGNs and normal Seyfert 1 galaxies, then similar 
variability should be observed in both systems with similar black holes. 
Indeed, those LLAGNs and AGNs follow the same trend in Figure~\ref{fig-1}, 
though LLAGNs have much lower luminosity at a given mass of the central black 
hole. (However, we should keep in mind that the variability of LLAGNs is 
possibly fundamentally different from that of normal AGNs.)
It can be predicted that many nearby LLAGNs should show little or no
significant short term variability because they may harbor ``dead'' QSOs
with larger black holes. This prediction is compatible with the fact that many
objects in Ptak et al. \shortcite{ptak} sample show little or no significant
short term variability.
 
Almaini et al. \shortcite{alm} found that the mean variability of QSOs in a
ROSAT sample
declines sharply with luminosity as seen in local AGNs, but with an upturn for
the most powerful sources. They argued that this is caused by evolution based
on the tentative evidence that the high redshift QSOs ($z>0.5$) do not show the
anti-correlation between variability and luminosity as seen in local AGNs. 
Based on our findings of the direct relation between the variability and
the mass of central black hole, a tentative interpretation is that the high
redshift QSOs is powered by a less massive black hole and is accreting more
efficiently than local AGNs at a given luminosity. 

It can be seen in Figure~\ref{fig-1} that the several NLS1s (NGC4051, MARK335
and PG1211+143) also follow the same trend with broad line objects.
Turner et al. \shortcite{turner} and Leighly \shortcite{leighly} found that
NLS1s have enhanced excess variability at a give luminosity. A simple 
explanation is that NLS1s have relatively
small black holes and enhanced accretion rates. Then the strong correlation
between  $\sigma^2_{\rm rms}$ and the FWHM of H$\beta$ can be consistently
explained \cite{turner}. Furthermore, Mathur \shortcite{mathur} recently
argued that NLS1s might be the Seyfert
galaxies and QSOs at an early stage of evolution and consequently have
small black holes at a given luminosity. He pointed out that they may be
low redshift analogues of the high redshift QSOs. This is consistent with
the observed fact that the high redshift QSOs have more variability
than local AGNs at a given luminosity \cite{alm}. If the relationship
between the variability and the mass of the central black hole is
really correct, it is intriguing that a systematic 
variability study for a large sample (covering a large luminosity and
redshift range) can give some information about the accretion history of
black holes and some constraints on the evolution scenario of
QSOs which has been debated for a long time. 

It should be cautioned that the ``excess variance'' of the NLS1s in our sample
is consistent with that of broad line objects rather than the NLS1s on the
plot of $\sigma^2_{\rm rms}$ versus the luminosity (see Figure~8 in Leighly
1999). One may ask about the other NLS1s. We noticed that MCG-6-30-15 is
also a NLS1 with a 1700km~s$^{-1}$ FWHM of H$\beta$ \cite{turner}, and it
does show enhanced excess variability on the plot of $\sigma^2_{\rm rms}$
versus luminosity (see the Figure 1 in Turner et al. 1999). Nowak \& Chiang
\shortcite{nch} recently studied its PDS, and found that the PDS is flat from 
approximately $10^{-6}-10^{-5}$Hz, and then steepens into a power law $\propto
f^{-\alpha}$ with $\alpha \ga 1$ and further steepens to $\alpha \approx 2$
between $10^{-4}-10^{-3}$Hz. Both the PDS shape and rms amplitude are
comparable to what has been observed in NGC5548 and CygX-1, though with break
frequencies differing by a factor of $10^{-2}$ and $10^4$. They pointed out
that the break frequency may indicate a central black hole mass as low as
$10^6$M$_{\sun}$. The monochromatic luminosity at 5100\AA of this object can
be approximately estimated to be $8\times10^{42}$ergs~s$^{-1}$
\cite{reynolds2}, and then its Virial mass can be determined by 
$ M=1.464\times10^5(R_{\rm BLR}/{\rm lt~days})(v_{\rm FWHM}/10^3{\rm km~s^{-1}})^2M_{\sun}\sim2\times 10^6M_{\sun}$,
by using the empirical law 
$R_{\rm BLR}=32.9^{+2.0}_{1.9}(\lambda L_{\lambda}(5100\AA)/10^{44}{\rm ergs~s^{-1}})^{0.700\pm0.023}{\rm lt~days}$ to estimate the location of broad line
region $R_{\rm BLR}$ (KSNMJG).
It gives a consistent central black hole mass with the one given by Nowak \&
Chiang \shortcite{nch}. Furthermore, this value is also consistent with the
one given by Reynolds et al. \shortcite{reynolds1} based on the time scale
arguments the 0.5-10keV flux increased by a factor of 1.5 over 100
seconds. We plotted this object in Figure~\ref{fig-1}, and found that it
consistently follows the same trend. Based on above discussion, it is possible
true that NLS1s show enhanced 
excess variation because they harbor smaller central black hole than normal
AGNs. However, further systematic investigations are needed to confirm the
above interpretation.

Some other models were also proposed to explain the trend of decreasing
variability with increasing luminosity. One of them is the multiple independent
hot spot model: the variability is caused by independent flaring events,
and more luminous systems simply have more flaring regions and hence smaller
$\sigma^2_{\rm rms}$. It cannot be simply ruled out since it is plausible
that a physical larger emission region will give rise to a greater number of
flares \cite{alm}. However, it is suspected that the flaring mechanism is the
same for both the objects with very lower $\dot{m}$ and those with higher
$\dot{m}$. In our sample, $\dot{m}$ must have a broad distribution: the
LLAGNs accrete material at a very lower rate and the luminous QSOs may accrete
material at high rate. Probably the X-ray emission mechanism for LLAGNs is
different from normal Seyfert 1 galaxies and QSOs. For example, the X-ray
may be emitted in an inner hot ADAF for LLAGNs, such as M81, NGC4258 and
NGC4579 \cite{gnb,quataert}, while it is possibly emitted from a corona for
normal Seyfert galaxies and QSOs. 
Another possibility is a toy
model of obscurational variability in AGNs, which can also give a relationship
between $\sigma^2_{\rm rms}$ and the size of central black hole, but in a
more complicated way and requiring fine tuning of several parameters, such as
the covering factor, the number of the clouds, etc. (see Abrassart \& Czerny
2000 for details).

The X-ray spectra of radio-loud objects are generally flatter than those
of radio-quiet objects. It is believed that the X-ray is emitted in the
jet rather than the disk-corona for radio-loud objects. So, the X-ray 
variability should be enhanced by relativistic effects. Three radio-loud
objects, 3C120, 3C390.3 and 3C373, are included in our sample. However,
they do not show significantly enhanced excess variability as shown
in Figure~\ref{fig-1}. This may be caused by that those objects have high
inclinations of the disk. Indeed, the best fit of the multi-band spectra of
3C273 give a high inclination of about $60^{\circ}$ \cite{kriss}. 

There are astrophysical reasons to relate the short-time X-ray variability
in AGNs with the accretion rate. For example, high variability can be
introduced by some increased instability in high accretion rate objects.
However, there are still no reliable estimation for the accretion rate
of AGNs. For example, the accretion rate of NGC4258, one of the best
studied object, is still in controversy. Its estimated values differ more
than one order of magnitude \cite{nm95,gnb}. There are separate large
amplitude flares in the light curves of many objects, especially in some
NLS1 possibly with high accretion rates; what's more, there exists some
flare events in the light curves of some NLS1s (e.g. PKS0558-504) which 
require a very large radiative efficiency (Gliozzi et al. 2000 and
references therein). Therefore, the short-time X-ray variability may be more
complicated than expected, and it may also depend on some other parameters.
A solid conclusion could be drawn by the current broad band X-ray mission like
Chandra and XMM-Newton.

\section{Conclusions}

To summarize, we have found a significant anti-correlation between the
``excess variance'' in X-ray band and the central black hole mass for a
composite sample of Seyfert 1 galaxies, QSOs and LLAGNs. Some simple global
coherent variations in the X-ray emission region, which scales with the
size of the central black hole, can lead to such a relation.
Based on our finding, the fact that NLS1 and high redshift QSOs show
enhanced excess X-ray variability than broad line local AGNs at a given
luminosity can be explained by that they harbor smaller black holes than
the normal local AGNs. These results highlight the significance of
pursuing a large-sample X-ray variability investigation, which can shed
light into the physical mechanism of AGNs and its evolution.

\section{Acknowledgments}

We thank an anonymous referee for helpful comments and suggestions
and Dr S. P. Oh for a careful reading of the manuscript.
YL acknowledges the hospitality of the Department of Astrophysical Sciences,
Princeton University.
This research has made use of data obtained through the High Energy 
Astrophysics Science Archive Research Center Online Service, provided by 
the NASA/Goddard Space Flight Center.

\end{document}